\documentstyle[11pt]{article}
\input{psfig}

\def\kmu{$K^+\to \mu^+ \pi^0 \nu$}

\def\kp{$K^+\to \pi^+ \pi^0$}

\def\rkm{$K^+\to \mu^+ \nu \gamma$}
\def\kppn{$K^+\to \pi^+ \pi^0 \pi^0$}

\begin{document}

\hspace{12cm} BNL-65154 

\begin{center}

{\Large \bf SEARCH FOR T-VIOLATION in $K_{\mu 3}$ DECAY}

{Presentation at the International Workshop on Kaon, Muon, 
and Neutrino Physics and the Future at KEK, October 31-November 1, 1997}

{Milind V. Diwan\footnote{For the E923 Collaboration:
 A. Carroll, M.V.~Diwan, J.~Frank, A.~Gordeev, S.~Kettell, L.~Leipuner, 
L.~Littenberg, H.~Ma, V.~Polychronakos (Brookhaven National Laboratory, 
Upton, NY, USA);
G.~Atoyan, V.~Issakov, O.~Karavichev, S.~Laptev, A.~Poblaguev, 
A.~Proskuryakov (Institute for Nuclear Research, Moscow, Russia);
 M.~Elaasar, D.~Greenwood, K.~Johnston (Louisiana Tech University, Ruston, 
LA,USA); R.~Adair , R.~Larson (Yale University, New Haven, CT, USA)} }

{Brookhaven National Laboratory, Upton, NY, USA}

{January 8, 1998}

\end{center}

\begin{abstract}
We have designed  a new experiment (E923) at the BNL-AGS to 
search for the T-violating polarization of
the muon normal to the decay plane of the $K^+ \to \mu^+ \pi^0 \nu$ decay.
The experiment aims to search for T-violation beyond the Standard Model; 
such a search is motivated by the need for a 
stronger CP violation source to account
for the baryon asymmetry of the Universe. 
 The experiment will be performed with in-flight decays from
 an intense ($2\times 10^7 K^+$/sec)
 2 GeV/c separated $K^+$ beam in an existing beam-line 
 at the AGS.
 We expect to analyze  more than $10^9$ events to
 obtain the sensitivity  of $\delta P_t = \pm 0.00013$
 at 1 $\sigma$, corresponding to the sensitivity of $\pm 0.0007$ to
$Im\xi$,
an improvement by 40 over the present limit on the same measurement.
\end{abstract}

\vspace{4ex}

\section*{Introduction}

Recently we examined the possibility of measuring various muon 
decay asymmetries that are sensitive to 
P, T or CP symmetries [\ref{ags2k}]; these are 
tabulated in Table \ref{list1}. 
Experimentally,  
CP violation has only been observed in  the neutral kaon system so far.
Although a  theoretical description of the 
CP-violation in the neutral kaon system exists through the complex phase 
in the Standard Model CKM matrix,  part or all of 
these phases could be consequences of 
deeper causes
that have so far eluded experiments.  
Over the last decade experiments 
at FNAL and CERN  directed towards the measurement of the direct 
$K^0_L \to \pi \pi$ transition or $\epsilon^\prime \over \epsilon$ have 
been inconclusive in  revealing the true nature of CP-violation.  Over 
the next decade ambitious efforts towards understanding CP-violation 
and the CKM matrix elements are planned with new $\epsilon^\prime
 \over \epsilon$
experiments and B-factories. The importance of these efforts is 
undeniable, yet
it must also be important to investigate the possibility that some or 
all of the 
CP-violation comes from effects outside the minimal Standard Model,  
particularly the CKM matrix.  

The CPT invariance of local quantum field theories requires
that CP violation is equivalent to T-violation. 
Therefore, it would be particularly interesting to look for 
direct violation of T-invariance outside the 
neutral kaon system.

It should also be noted that  CP-violation is required to generate  
the observed baryon asymmetry of the universe, and it 
is now accepted that  the CP-violation embodied in the CKM 
matrix does not have sufficient strength for this purpose [\ref{mclerran}]. 
Physics beyond the Standard Model that could generate the baryon 
asymmetry   can also generate 
CP or T violating muon polarizations in the kaon decay modes in Table \ref{list1}. 

Muon polarizations from kaon decays have a rich phenomenology.
In the case of $K_L \to \mu^+ \mu^-$ and 
$K^+ \to \pi^+ \mu^+ \mu^-$
new measurements could lead to important constraints on 
the Standard Model CKM parameters, in particular 
the Wolfenstein parameters $\rho$ and $\eta$. 
It is, however, difficult to reach the level of sensitivity needed 
to measure these parameters well with current technology. Nevertheless,
the experimental  difficulties should be compared to the difficulties
facing the rare kaon decay measurement of $K_L \to \pi^0 \nu \bar\nu$,
which   is sensitive to the same physics.

As shown in Tables 
\ref{list1} and \ref{list2}  
for many cases 
limits on the muon polarization will probe new physics 
beyond the Standard Model. In particular, 
the polarization will be sensitive to the physics of
a more complicated Higgs sector or leptoquarks that could 
give rise to CP or T violation outside the Standard Model.
The other source of CP violation needed for baryogenesis could be  
the motivation for such searches. In the case of $K^+\to \pi^0 \mu^+ \nu$ 
and $K^+\to \mu^+ \nu \gamma$ decays large gains in the sensitivity to 
amplitudes not allowed in the Standard Model are possible with current 
detector techniques and available beams. Therefore we have designed 
the new experiment to  focus on  these measurements.

\begin{table}
\begin{center}
\begin{tabular}{clccc}
\hline
\hline
 & Decay                &  Correlations & Symmetries \\
  &                     &              &  tested    \\
\hline
(1) & $K^+\to \pi^0 \mu^+ \nu$   &  $\vec s_\mu\cdot (\vec p_\mu\times \vec p_\pi)$ & T \\
\hline
(2) & $K^+\to  \mu^+ \nu \gamma$  &  $\vec s_\mu\cdot (\vec p_\mu\times \vec p_\gamma)$ & T \\
\hline
(3) &  $K_L\to  \mu^+ \mu^-$ &   $\vec s_\mu\cdot \vec p_\mu$ & P, CP \\
\hline
(4) &  $K^+\to \pi^+ \mu^+ \mu^-$ &  $\vec s_\mu\cdot \vec p_\mu$ & P \\
(5) &                             & $\vec s_\mu\cdot (\vec p_{\mu^+}\times \vec p_{\mu^-})$ & T \\
(6) &                             & ($\vec s_\mu \cdot \vec p_\mu) \vec s_\mu\cdot (\vec p_{\mu^+}\times \vec p_{\mu^-})$ & P, T \\
\hline
\hline 
\end{tabular}
\end{center}
\caption{\sl The decay modes and the polarization asymmetries 
or correlations of interest. }
\label{list1}
\end{table}

\begin{table}
\begin{center}
\begin{tabular}{llccccc}
\hline
\hline
Asym. & Mode & Branch. &  Standard   & Final      &  Non-SM    & Ref. \\
      &      & Fraction  &  Model    & State Int. &  value      &      \\
\hline 
(1) & $K^+\to \pi^0 \mu^+ \nu$ & 0.032 & 0.0 & $\sim 10^{-6}$ & $\le 10^{-3}$ & [\ref{garisto}] \\
\hline
(2) & $K^+\to \mu^+ \nu \gamma$ & $5\times 10^{-3}$ & 0.0 & $\sim 10^{-3}$ & $\le 10^{-3}$ & [\ref{kobayashi}] \\
\hline
(3) & $K_L \to \mu^+ \mu^-$ & $7\times 10^{-9}$ & $\sim 10^{-4}$ & 0.0 & $\le 10^{-2}$ & [\ref{gengng}, \ref{wolf1}] \\
\hline
(4) &   $K^+\to \pi^+ \mu^+ \mu-$ & $5\times 10^{-8}$  &  $\sim 10^{-2}$ & -- & -- & [\ref{wise1} --\ref{bgt}] \\
(5) &                &  & 0.0 & $\sim 10^{-3}$  & $\sim 10^{-3}$ & [\ref{anbg},\ref{gengsum}] \\
(6) &               &   & $\sim 6\times 10^{-2}$ & $\sim 0.0$ & $\sim 0.1$ & [\ref{anbg},\ref{gengsum}] \\
\hline
\end{tabular}
\end{center}
\caption{\sl The decay modes and asymmetries discussed by the working group.
The asymmetries are numbered as in Table \ref{list1}. 
The rest of the columns are:
the known branching ratio,
 the estimated Standard Model value,
the value due to final state interactions,
 the maximum possible
value allowed by non-standard  
physics, and the theoretical reference. }
\label{list2}
\end{table}

\section*{$K^+\to \pi^0 \mu^+ \nu$} 
The transverse or  out of plane muon polarization  ($P^\mu_T(K_{\mu3})$) in this 
 decay has been analyzed by many authors [\ref{garisto},
\ref{belanger}, \ref{gengsum}]. The out of plane polarization is
expected to be zero to first order 
in the Standard Model because of the absence of 
the CKM phase in the decay amplitude. It has been shown that any 
arbitrary models involving effective V or A interactions cannot 
produce this type of polarization. 
 Irreducible  
backgrounds, i.e., final state interactions (FSI), to the 
out of plane polarization in this decay are expected to be 
small ($\sim 10^{-6}$) and therefore can be ignored [\ref{zitnitskii}].
Therefore, the existence of a non-zero
value of this polarization will be a definite signature of new 
physics.  
In particular, some multi-Higgs  and 
leptoquark models could produce such a polarization. In multi-Higgs
models  a charged Higgs particle 
mediates an effective scalar interaction that interferes 
with the Standard Model decay amplitude; in such models 
the polarization could be as large as $10^{-3}$ without 
conflicting with other experimental constraints including 
the measurements of the neutron electric dipole moment and 
the branching fraction for $B\to X \tau \nu$, or $b\to s \gamma$ 
[\ref{ksmith}-\ref{aleph2}]. In Table \ref{tab_of_con} we have listed 
many such constraints and translated them into a limit on the T-violating
polarization in $K_{\mu3}$ decays. As can be seen the T-violating muon
polarization is the best probe of this physics. The out of plane 
polarization of $\tau$'s in $B\to D \tau \nu$ decays has recently been examined
in the same context [\ref{kiers}]. Depending on the phase space examined a factor 
of 30 enhancement of the lepton polarization is possible in B semileptonic decays 
to $\tau$ relative to \kmu ~decay. Therefore a 
measurement of the $\tau$ out of plane polarization within $\pm 0.4\%$ is needed to match the 
sensitivity that we have proposed for \kmu ~decays. Such a measurement with
B decays needs high statistics and is possible with the proposed 
next generation B-factories if
 the $\tau$ polarization can be analyzed with high efficiency.


\begin{table}
\begin{tabular}{|l|l|l|}
\hline
Measurement   &   Value (or $95\%$ C.L.)   &   $\bar P^T_\mu < $  \\
               &         &     $m_h\sim 2 m_W$ \\ 
\hline 
$d_n$ [\ref{ksmith}] & $< 1.2\times 10^{-25} e-cm$ (95\%C.L.)&  $<0.064$ \\
$d_e$  [\ref{abdullah}] & $< 1.7\times 10^{-26} e-cm$ (95\%C.L.)&  -- \\
$\epsilon^\prime\over \epsilon$ [\ref{eprime}] & $(1.5\pm0.8)\times 10^{-3}$ & -- \\
$m_{K_L} - m_{K_S}$ [\ref{mklks}] & $(3.510\pm0.018)\times 10^{-6} eV$ &  -- \\
$B(b\to s ~\gamma)$ [\ref{alam}] & $(2.32\pm 0.67)\times 10^{-4}$ &  $<0.02$ \\
$P^\mu_T(K_{\mu3})$ [\ref{campbell}] &               & $-0.009<P^T_\mu <0.007$  \\
$B(b\to X \tau \nu)$ [\ref{aleph2}] & $(2.72\pm 0.34)\times 10^{-2}$ & $<0.008$  \\  
\hline
\end{tabular}
\caption{ \sl Constraints on $ P^T_\mu$ for 3HDM [\protect\ref{weinberg}] from various measurements. 
For details see Ref.  [\protect\ref{garisto}] and [\protect\ref{belanger}].
Also see Ref. [\protect\ref{kuno2}].
We have assumed that the 3 vacuum expectation values have the same 
ratios as the third generation masses;  $v_1:v_2:v_3::m_b:m_t:m_\tau$; 
and $m_t\sim 170 GeV$ and $m_h\sim 2m_W$.
(--) means that there is no significant constraint.
All constraints obtained are at 95\% confidence level. }
\label{tab_of_con}
\end{table}

The best previous experimental limits were obtained  almost 15 years 
ago with both 
neutral [\ref{schmidt}] and charged kaons [\ref{campbell}] at the BNL-AGS.
 The experiment with $K^+$ decays produced 
a measurement of the transverse polarization, $P^T_\mu = 0.0031 \pm 0.0053$.
The combination of both experiments could be interpreted as a limit on the 
 imaginary part of the ratio of the hadron form factors,
$Im\xi = Im(f_-/f_+) = -0.01 \pm 0.019$.
This limit is mostly independent of theoretical models 
and the experimental acceptance. 
By using the approximate formula, 
$P^T_\mu \approx 0.183 \times Im\xi$, one may reinterpret the above measure
of $Im\xi$ as a combined limit on the polarization, $P^T_\mu \approx 
-0.00185 \pm 0.0036$.  
This 1980 era measurement was based on  
$1.2\times 10^7 ~K^0_L$ 
and $2.1\times 10^7 ~ K^+$ decays to $\mu^+ \pi  \nu$ and was 
limited by statistics and backgrounds.  

Currently an experiment is in progress at the KEK-PS, E246 [\ref{kek246}], 
to measure $ P^T_\mu$ with a new technique of using 
a stopping $K^+$ beam and 
measuring the muon decay direction 
without spin precession. 
They expect to 
reach a sensitivity of $9\times 10^{-4}$ ($Im \xi < 4\times 10^{-3}$)
with $1.8\times 10^7$ events. 
The experiment will try to minimize systematics
by using the cylindrical symmetry of the apparatus and by using the
backward-forward $\pi^0$ 
symmetries of the decay at rest. 
The results of this
experiment will be very valuable to future experiments. Preliminary 
results from  their
recent data are reported elsewhere in this proceedings.

\section*{$K^+ \to \mu^+ \nu \gamma$}
The T violating out of plane polarization of the muon 
in this decay also probes non Standard Model physics    
similar  to the  $K^+\to \pi^0 \mu^+ \nu$ decay. The former 
can be caused by an effective pseudo-scalar interaction, while 
the latter  by an effective scalar interaction. 
In addition, it is also sensitive to non-Standard Model vector
and axial vector couplings. 
Therefore searches for T violation  
in both decay modes are complementary [\ref{kobayashi}].  
The T violating polarization in \rkm ~could be  $\sim 10^{-3}$ 
without violating other experimental bounds. It is estimated [\ref{marciano},\ref{gengrkm2}] 
that  the electro-magnetic 
FSI for this interaction can induce an out of plane muon polarization
of the same order of magnitude. An accurate theoretical calculation will 
be needed to subtract the FSI from any observation.
On the other hand, this FSI induced effect 
could be considered a useful calibration point for the apparatus that will 
also be used for the new $K^+ \to \pi^0 \mu^+ \nu$ experiment.

\section*{Detector Design}

\begin{figure}
\begin{center}
\psfig{figure=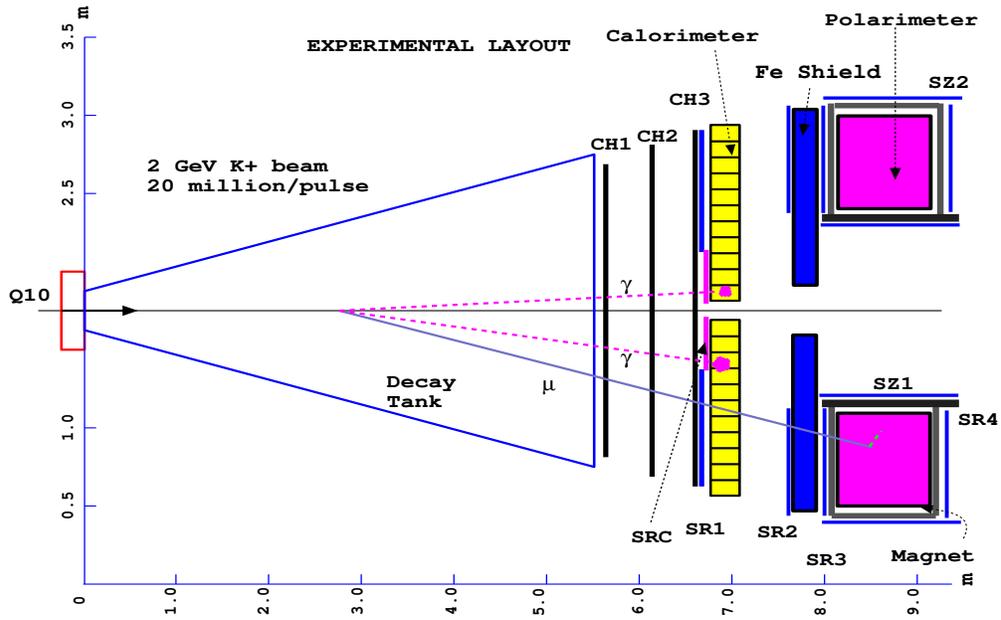,height=3.2in,width=5.2in,angle=90} 
\caption{Schematic of the detector.  A typical $K^+\rightarrow
\mu^+\pi^0\nu$ events is superimposed.}
\end{center}
\label{pict1}
\end{figure}
The E923 experiment will be performed with 2 GeV/c electro-statically
separated charged kaons
decaying in flight. 
The beam intensity will be $2\times 10^7 K^+$'s/spill with 
$3\times 10^{13}$ protons on target every 3.6 sec. 
Figure \ref{pict1}
 shows the plan  view
 of the experiment.  The basic workings of the experiment
 are the same as the experiment in Ref. [\ref{campbell}].
 The detailed design is, however, optimized for a
 high intensity 2 GeV beam. 
The cylindrically symmetric detector
is centered on the kaon beam. The $K^+_{\mu 3}$ ~decays of interest
occur in the decay tank;  the photons from the
$\pi^0$ decay are detected in the calorimeter; the muon stops
in the polarimeter.
The decay of the stopped muon is 
detected in the polarimeter by wire chambers, which are
arranged radially with 96 graphite wedges that serve as absorber medium.
The hit pattern in the polarimeter identifies the muon stop as well as 
positron direction relative to the muon stop.
By selecting events with $\pi^0$ moving along the beam direction and
muon moving perpendicular to the beam direction in the $K^+$ center of  
mass frame, the decay plane coincides with the radial wedges. 
A  non-zero transverse muon polarization causes an asymmetry between
the number of muons that decay
clockwise versus the number counter-clockwise.
To reduce systematic errors, a weak solenoidal magnetic field along
the beam direction 
(70 gauss or a precession period of $\sim 1 \mu s$) with
polarity reversal every spill is applied to the polarimeter. 
If there is an initial muon transverse polarization, there will be
 a small shift in the
phase of the sinusoidal oscillation in the measured asymmetry. 
The difference in the asymmetry for the two polarities is proportional 
to the T-conserving  muon polarization in the decay plane, while the sum is
proportional to the T-violating muon polarization normal to the decay plane. 
Both components will have the same frequency but will be 90 degrees apart in 
phase. 

Compared to the previous in-flight experiment,  this experiment has
much better background rejection and event reconstruction.  
The separated $K^+$ beam should greatly reduce the accidental rate.
The polarimeter is finely segmented and the analyzing power is higher. 
The positron signature is defined by the coincidence of signals
in a pair of neighboring wedges.
The larger calorimeter makes it possible to 
measure both photons and to 
reconstruct the $\pi^0$ 
momentum
for a large fraction of the events.
 Together with the muon trajectory, the events can be fully
reconstructed.  
The detector
acceptance and background rejection is optimized using GEANT
simulation.

The experiment will collect approximately 550 events per AGS 
pulse per 3.6 seconds. Thus the statistical accuracy of the 
polarization measurement in a 2000 hr ($2\times 10^6$ pulses) run 
will be:
\begin{eqnarray}
\delta P_T \approx { 1.20^{1\over 2} 2^{1\over 2} \over 0.35
             (2\times 10^6 \cdot 550)^{1\over 2}  } 
            \approx 1.3\times 10^{-4} \nonumber
\end{eqnarray}
where $\sqrt{1.2} $,  $\sqrt 2$, $0.35$, are dilution factors in the 
analyzing power due to backgrounds, the precession magnetic field, 
and the muon 
decay, respectively.  The sensitivity to $Im{\xi}$ is given by 
$$\delta Im{\xi} \approx \delta P_T/0.2 \approx 7\times 10^{-4}$$
where 0.2 is a kinematic factor that includes the acceptance in 
the Dalitz plot and the orientation of the decay in the 
center of mass.

This new experiment  is optimized to study muon polarization in \kmu 
~decays.   Nevertheless, we have investigated the 
feasibility of measuring T-violation in \rkm.
The event selection and analysis of \rkm ~will be 
very similar to \kmu ~events except 
that events containing more than 1 photon 
 will be vetoed to reject background 
from \kmu, \kp, and \kppn ~events.
Further background rejection will be 
achieved by matching the measured muon 
 range in the 
polarimeter with the muon energy from a constrained fit to
the photon momentum, the muon direction, and the known 
kaon momentum. 
We expect to collect  $\sim 100$
events per AGS pulse. 
However,  the signal to background ratio  
with out current design will be 
about  0.3.
Two improvements to the detector will reduce the backgrounds 
further:  
If the decay volume can be surrounded by photon veto counters with 
a veto threshold of 10 MeV
to detect the low energy photons from $\pi^0$ decays,
 the background level can be reduced to 
about 10\%. Secondly, if the calorimeter resolution can be improved 
(we have assumed $\sigma(E)/E \sim 8\%/\sqrt{E}$) then the 
muon range match can be made narrower, thus separating the
signal and background better. We are currently calculating the 
sensitivity that can be gained with  these modest improvements 
 for the 
transverse muon polarization in \rkm.

\section*{Systematic Errors}

With such high statistical power, systematic issues will become 
the main concern. The cylindrical symmetry of the apparatus 
and the precession technique (see [\ref{campbell}])
will cancel most systematic errors to first order. 
We have identified several sources of systematic error
and are working on strategies to either eliminate or calibrate 
them with data. Angular
misalignment of the polarimeter wedges will need to be controlled by aligning
the opposite side wedges with respect to each other.
There could be systematic shifts in detector time calibration which depend
on the sign of the magnetic field that precesses the muons in the 
polarimeter. 
Stray magnetic fields could precess the muons 
before they come to rest or alter the magnitude of the precession field
inside the polarimeter. We are currently working on techniques to 
precisely align the detector and effectively  shield stray magnetic fields.
Other sources of systematic errors are beam and calorimeter misalignments
combined with non-uniform efficiencies as well as momentum dispersion 
within the beam. We have found that these sources contribute at second order
and can easily be reduced further with proper analysis techniques. 

The experiment will collect a large sample of data including 
$K^+ \to \mu^+ \nu$, $K^+ \to \pi^+ \pi^0$, and $K^+\to \pi^0 \mu^+ \nu$
events in different parts of the decay phase space. The muon decay 
asymmetries from these various ensembles of events can be measured 
to understand the detector systematics to very high accuracy. 
We will use the T-conserving component of the muon polarization to calibrate
the analyzing power throughout the detector.

\section*{Conclusion}

We have examined the measurement of the out of plane 
muon polarization in \kmu  ~decays in detail. This 
 measurement will not be sensitive to the Standard Model 
CP violation physics. 
Nevertheless, it  can be 
performed with great sensitivity;
 approaching $\delta P \sim 10^{-4}$ 
which is well beyond both the current direct limit 
of $\sim 5.3\times 10^{-3}$.
Although the electric dipole moments  of the
neutron and electron are considered more favorably
for T violation outside the Standard Model they do not 
cover the entire  spectrum of models. 
At the moment the measurement of T violating 
polarization in \kmu ~decays 
is well justified and should be considered complementary 
to other efforts in understanding CP violation.
We are also examining 
the sensitivity for transverse polarization 
measurement in \rkm ~decays 
after some improvements to the present design of E923.
There is presently no direct limit on such a measurement, and therefore
a sensitive experiment to measure this polarization will be 
quite interesting and complementary to the \kmu ~measurement.  
The experiment E923 has been approved by the Brookhaven National Laboratory. 
We are currently developing the  detectors and the 
techniques to achieve the required 
level of systematic control. 

This work was supported by US-DOE contract DE-AC02-76CH00016.

\bigskip
\bigskip

{\bf REFERENCES}

\begin{enumerate}

\vspace{-10pt}
\item \label{ags2k}
``Muon Polarization Working Group Report'', R. Adair et al., 
Workshop on AGS experiments for the 21st Century, May 12-17, 1996.
\vspace{-10pt}
\item \label{mclerran}
 L. Mclerran,  M. Shaposhnikov, N. Turok, and M. Voloshin, Phys. Lett. 
{\bf B 256}, 
451 (1991).  N. Turok and M. Voloshin, Phys. Lett. {\bf B 256}, 451 (1991). 
N. Turok and J. Zadrozny, Nucl. Phys. {\bf B 358}, 471 (1991). M. Dine, 
P. Huet,
R. Singleton, and L. Susskind, Phys. Lett. {\bf B 257}, 351 (1991). 
\vspace{-10pt} 
\item \label{garisto}
 R. Garisto and G. Kane, Phys. Rev. {\bf D 44}, 2038 (1991).
\vspace{-10pt}
\item \label{belanger}
 G.  Belanger and C. Q. Geng, Phys. Rev. {\bf D 44}, 2789 (1991).
\vspace{-10pt}
\item \label{gengsum}
C.Q. Geng, Talk presented at the KEK workshop on rare kaon 
decays, UdeM-LPN-TH-79, Dec. 10, 1991. 
\vspace{-10pt}
\item \label{zitnitskii}
A. R. Zhitnitskii, Sov. J. Nucl. Phys. {\bf 31}, 529 (1980).
\vspace{-10pt}
\item \label{ksmith}
K. Smith, et al., Phys. Lett. B {\bf 234}, 191 (1990).
I.S. Altarev, et al., Phys. Lett. B {\bf 276}, 242 (1992).
\vspace{-10pt}
\item \label{abdullah}
K. Abdullah, et al., Phys. Rev. Lett. {\bf 65}, 2347 (1990). 
\vspace{-10pt}
\item \label{eprime}
G.D. Barr, et al., Phys. Lett. B{\bf 317}, 233 (1993).
L.K. Gibbons, et al., Phys. Rev. Lett. {\bf 70}, 1203 (1993).
The average value quoted by the PDG is used from
 Phys. Rev. D {\bf 50}, 1545 (1994).
\vspace{-10pt}
\item \label{mklks} 
L.K. Gibbons, et al., Phys. Rev. Lett. {\bf 70}, 1199, (1993).
Particle Data Group, Phys. Rev. D {\bf 50}, 1545 (1994).
\vspace{-10pt}
\item \label{alam} 
M.S. Alam, et al., Phys. Rev. Lett. {\bf 74}, 2885 (1995)
For the theoretical treatment in the context of the 3 Higgs
doublet model (3HDM) [\ref{weinberg}] see
Y. Grossman and Y. Nir, Phys. Lett. B {\bf 313}, 126 (1993).
\vspace{-10pt}
\item \label{aleph2} 
The ALEPH Collaboration, Paper Contributed to the International Conference on 
High Energy Physics, Warsaw, Poland, 25-31 July 1996. PA10-019. \\
http://alephwww.cern.ch/ALPUB/oldconf. \\
For the theoretical treatment in the context of 3HDM see
Y. Grossman, 
Nuclear Physics {\bf B426}, 355 (1994).
\vspace{-10pt}
\item \label{kiers}
G.H. Wu, Ken Kiers, J,N, Ng, Phys. Rev. {\bf D56}, 5413 (1997).
\vspace{-10pt}
\item \label{schmidt}
 M. Schmidt, et al., Phys. Rev. Lett. {\bf 43}, 556 (1979).  W. Morse, et al., 
Phys. Rev.  {\bf D 21}, 1750 (1980).
\vspace{-10pt}
\item\label{campbell}
 M. Campbell, et al.,
 Phys. Rev. Lett. {\bf 47}, 1032 (1981). S. Blatt, et al., 
Phys. Rev.  {\bf D 27}, 1056 (1983).
\vspace{-10pt}
\item \label{kek246} 
J. Imazato, et al., KEK-PS research proposal Exp-246,
June 6, 1991.
\vspace{-10pt}
\item \label{kuno2}
Y. Kuno, Nucl. Phys. Proc. Suppl. {bf 37 A}, 87 (1993).
3rd KEK Topical Conference on CP Violation, Its Implications to Particle 
Physics and Cosmology, KEK, Tsukuba, Japan, Nov 16-18, 1993.
\vspace{-10pt}
\item \label{weinberg}
 S. Weinberg, Phys. Rev. Lett. {\bf 37}, 657 (1976).
\vspace{-10pt}
\item \label{kobayashi}
M. Kobayashi, T.-T. Lin and Y. Okada, Progress of Theoretical Physics, 
{\bf 95}, 261 (1996).
\vspace{-10pt}
\item \label{marciano}
W. Marciano, Private Communication.
See also C.Q. Geng, Nucl. Phys. B (Proc. Suppl.) {\bf 37A}
59 (1994). 
\vspace{-10pt}
\item \label{gengrkm2}
C.Q.Geng, Nucl. Phys. B {\bf 37A}, 59 (1994).
\vspace{-10pt}
\item \label{gengng}
F.J. Botella and C.S. Lim, Phys. Rev. Lett. 56 (1986) 1651.
Also see
C.Q. Geng, J.N. Ng, TRI-PP-90-64, Paper presented at the 
BNL CP summer study,
May 21-22, 1990.
\vspace{-10pt}
\item \label{wolf1}
J. Liu, L. Wolfenstein, Nuclear Physics, {\bf B289} 1 (1987).
\vspace{-10pt}
\item \label{wise1} 
Ming Lu, Mark B. Wise, and Martin J. Savage, Phys. Rev. {\bf D46} 
5026 (1992).
\vspace{-10pt}
\item \label{wise2}
Martin J. Savage, Mark B. Wise, Phys. Lett. {\bf B250} 151 (1990).
\vspace{-10pt}
\item \label{buchalla}
G. Buchalla, A.J. Buras,  Phys. Lett. {\bf B336} 263 (1994).
\vspace{-10pt}
\item \label{gourdin}
Michel Gourdin, PAR-LPTHE-93-24, May 1993. 
\vspace{-10pt}
\item \label{kuno1}
 Yoshitaka Kuno (KEK, Tsukuba). 
KEK-PREPRINT-92-190, Jan 1993. 4pp. Published in
Proc. 10th Int. Symp. on High Energy Spin Physics, Nagoya, Japan, 
Nov 9-14, 1992. Page 769. 
\vspace{-10pt}
\item \label{bgt}
By G. Belanger, C.Q. Geng, P. Turcotte, 
Nucl. Phys. {\bf B390} 253 (1993).
\vspace{-10pt}
\item \label{anbg}
 Pankaj Agrawal, John N. Ng, G. Belanger, C.Q. Geng,
Phys. Rev. {\bf D45} 2383 (1992).
\vspace{-10pt}

\end{enumerate}

\end{document}